%% file: VaMoS2020.tex
\documentclass[sigconf, authorversion=true]{acmart} 
\input{docSettings}
\input{lstSettings}

\usepackage[all=normal, paragraphs=tight]{savetrees}

\setcopyright{acmlicensed}
\copyrightyear{2020}
\acmYear{2020} 
\acmConference[VaMoS '20]{Proceedings of the 14th International Working Conference on Variability Modelling of Software-Intensive Systems}{February 5--7, 2020}{Magdeburg, Germany}
\acmBooktitle{Proceedings of the 14th International Working Conference on Variability Modelling of Software-Intensive Systems (VaMoS '20), February 5--7, 2020, Magdeburg, Germany}
\acmPrice{15.00}
\acmDOI{10.1145/3377024.3377031}
\acmISBN{978-1-4503-7501-6/20/02}

\begin{document}
\input{000-frontmatter}
\input{010-introduction}
\input{020-relatedWork}
\input{030-Tradeoffs}
\input{040-concept}
  \input{042-RAST}
  \input{043-application}
\input{050-realization}
\input{070-benchmark}
\input{060-discussion}
\input{080-conclusion}


\vfill
\begin{acks}
This work is partially supported by the ITEA3 project $\text{REVaMP}^2$, funded by the \grantsponsor{01IS16042H}{BMBF (German Ministry of Research and Education)}{https://www.bmbf.de/} under grant \grantnum{01IS16042H}{01IS16042H}. Any opinions expressed herein are solely by the authors and not of the BMBF.
\end{acks}

\clearpage
\bibliographystyle{ACM-Reference-Format}
\balance
\bibliography{literature}

\end{document}

%% file: docSettings.tex
\usepackage[english]{babel}
\usepackage[utf8]{inputenc}
\usepackage[T1]{fontenc}

\usepackage{amsmath}

\usepackage{multirow}

\usepackage{graphicx}

\usepackage{titlesec}
\titlespacing*{\section}
	{0pt}                     
	{1.2ex plus 0ex minus .4ex} 
	{.5ex}                    
\PassOptionsToPackage{hyphens}{url}
\usepackage{hyperref}
\hypersetup{
    pdfauthor={Sascha El-Sharkawy, Adam Krafczyk, and Klaus Schmid},
    pdftitle={Fast Static Analyses of Software Product Lines - An Example With More Than 42,000 Metrics},
    pdfsubject={14th International Working Conference on Variability Modelling of Software-Intensive Systems (VaMoS '20)},
    pdfkeywords={Software Product Lines; SPL; Metrics; Implementation; Variability Models; Feature Models; Abstract Syntax Tree; AST},
    pdfpagemode={UseOutlines},
    pdfproducer={LaTeX},
    colorlinks=true,
    linkcolor=black,          
    citecolor=black,          
    filecolor=black,          
    urlcolor=black            
}

\usepackage{enumitem}
\setlist[itemize]{leftmargin=1em}
\setlist[enumerate]{leftmargin=3ex}
\setlist[description]{leftmargin=1em}
\newcommand{\RQ}[1]{\textbf{RQ#1}}
\newcommand{\RQLayerOneLabel}{\RQ{\arabic*}}
\newcommand{\ReqLabel}{\textbf{Req\arabic*}}

\usepackage{flushend}

\usepackage{pifont}
\newcommand{\cMark}{\ding{51}}          
\newcommand{\partialSupport}{\ding{108}}
\newcommand{\noSupport}{\ding{55}}

\newcommand{\absTitle}[1]{\textit{#1:}}

\newcommand{\ifDef}{\texttt{\#ifdef}}
\newcommand{\elseCPP}{\texttt{\#else}}
\newcommand{\elif}{\texttt{\#elif}}
\newcommand{\ifDefBlock}{\ifDef-block}

\newcommand{\kernelhaven}{\texttt{KernelHaven}}
\newcommand{\srcml}{\texttt{srcML}}
\newcommand{\srcmlextractor}{\texttt{srcMLExtractor}}
\newcommand{\kbuildminerextractor}{\texttt{KbuildMiner\-Extractor}}
\newcommand{\kconfigreaderextractor}{\texttt{KconfigReader\-Extractor}}
\newcommand{\metrichaven}{\texttt{MetricHaven}}

\newcommand\footnoteref[1]{\protected@xdef\@thefnmark{\ref{#1}}\@footnotemark}

\newcommand{\SD}{SD}
\newcommand{\SDvp}{\SD$_\text{VP}$}
\newcommand{\SDfile}{\SD$_\text{File}$}

\makeatletter
  \newcommand{\thickhline}{%
      \noalign {\ifnum 0=`}\fi \hrule height 1pt
      \futurelet \reserved@a \@xhline
  }
\usepackage{array}
\newcolumntype{"}{@{\hskip\tabcolsep\vrule width 1pt\hskip\tabcolsep}}
\makeatother

\newcommand{\tRef}[1]{\tikz[remember picture]\node[](#1){};\hspace{-5pt}}
\usepackage{tikz}
\usetikzlibrary{matrix,arrows,positioning,shadows}
\usetikzlibrary{shapes.arrows,fit,shapes.geometric}
\usetikzlibrary{chains}
\tikzset{%
  block/.style = {draw=red!70!black, thick, rectangle},
}

\usepackage{environ}
\makeatletter
	\newsavebox{\measure@tikzpicture}
	\NewEnviron{scaletikzpicturetowidth}[1]{%
		\def\tikz@width{#1}%
		\begin{lrbox}{\measure@tikzpicture}%
		\BODY
		\end{lrbox}%
		\pgfmathparse{#1/\wd\measure@tikzpicture}%
		\BODY
	}
\makeatother

\usepackage{adjustbox}

%% file: lstSettings.tex
\definecolor{cEclipseRed}{rgb}{0.6,0,0}
\definecolor{cEclipseGreen}{rgb}{0.25,0.5,0.35}
\definecolor{cEclipsePurple}{rgb}{0.5,0,0.35}
\definecolor{cEclipseBlue}{rgb}{0.25,0.35,0.75}
\definecolor{ifDefRed}{rgb}{0.753,0,0}

\usepackage{listings}
\lstdefinelanguage{config} {
  language=C,
  morecomment=[l]{\#},
  morekeywords=[2]{y, n, m},
  keywordstyle=[2]{\color{red!70!black}}
}

\lstdefinelanguage{kconfig} {
  language=config,
  string=[b]",
  keywords = {choice, endchoice, config, bool, tristate, string, int, hex, select, depends, on, option, modules, default, def_bool, def_tristate, range, if, endif, menu, endmenu, prompt, source, help}
}

\lstdefinelanguage{kbuild} {
  language=[gnu]make,
	moredelim=[is][stringstyle]{@<}{>@}
}

\lstdefinelanguage{cWithPre} {
  language=C,
  morecomment=[l][\bfseries\color{ifDefRed}]{\#},
}

\lstdefinelanguage{pseudo} {
    language=Java,
    morekeywords=[2]{end, done, begin},
    mathescape
}

\lstdefinelanguage{properties} {
    tabsize=1,
    commentstyle=\color{blue}, 
    morecomment=[s][\color{cEclipsePurple}]{<}{>},
		morecomment=[l]{\	},
    morecomment=[l][\color{cEclipseGreen}]{\#},		
}

%% file: 000-frontmatter.tex
\title[Fast Static Analyses of Software Product Lines --- An Example With More Than 42,000 Metrics]{Fast Static Analyses of Software Product Lines\texorpdfstring{\\}{ }--- An Example With More Than 42,000 Metrics}

\author{Sascha El-Sharkawy}
\affiliation{%
  \institution{University of Hildesheim, Institute of Computer Science}
  \streetaddress{Universit{\"a}tsplatz 1}
  \city{Hildesheim}
	\country{Germany}
  \postcode{31134}
}
\email{elscha@sse.uni-hildesheim.de}

\author{Adam Krafczyk}
\affiliation{%
  \institution{University of Hildesheim, Institute of Computer Science}
  \streetaddress{Universit{\"a}tsplatz 1}
  \city{Hildesheim}
	\country{Germany}
  \postcode{31134}
}
\email{adam@sse.uni-hildesheim.de}

\author{Klaus Schmid}
\affiliation{%
  \institution{University of Hildesheim, Institute of Computer Science}
  \streetaddress{Universit{\"a}tsplatz 1}
  \city{Hildesheim}
	\country{Germany}
  \postcode{31134}
}
\email{schmid@sse.uni-hildesheim.de}

\begin{abstract}
\absTitle{Context}
Software metrics, as one form of static analyses, is a commonly used approach in software engineering in order to understand the state of a software system, in particular to identify potential areas prone to defects. Family-based techniques extract variability information from code artifacts in Software Product Lines (SPLs) to perform static analysis for all available variants.
Many different types of metrics with numerous variants have been defined in literature.
When counting all metrics including such variants, easily thousands of metrics can be defined.
Computing all of them for large product lines can be an extremely expensive process in terms of performance and resource consumption.

\absTitle{Objective}
We address these performance and resource challenges while supporting customizable metric suites, which allow running both, single system and \textit{variability-aware} code metrics.

\absTitle{Method} 
In this paper, we introduce a partial parsing approach used for the efficient measurement of more than 42,000 code metric variations. The approach covers variability information and restricts parsing to the relevant parts of the Abstract Syntax Tree (AST).

\absTitle{Conclusions}
This partial parsing approach is designed to cover all relevant information to compute a broad variety of variability-aware code metrics on code artifacts containing annotation-based variability, e.g., realized with C-preprocessor statements. It allows for the flexible combination of single system and variability-aware metrics, which is not supported by existing tools. This is achieved by a novel representation of partially parsed product line code artifacts, which is tailored to the computation of the metrics. Our approach consumes considerably less resources, especially when computing many metric variants in parallel.
\end{abstract}

%
%

\ccsdesc[500]{General and reference~Metrics}
\ccsdesc[500]{Software and its engineering~Software product lines}
\ccsdesc[300]{Software and its engineering~Automated static analysis}

\keywords{Software Product Lines, SPL, Metrics, Implementation, Variability Models, Feature Models, Abstract Syntax Trees, AST}
\maketitle

%% file: 010-introduction.tex
\section{Introduction}
\label{sec:Introduction}
In software engineering, static analyses are commonly used in order to analyze a software system and to identify potential defects. A well established form of static analyses are software metrics \cite{FentonBieman14}, which are used for the prediction of faults \cite{RadjenovicHerickoTorkar+13} or maintainability issues \cite{RiazMendesTempero09}. 
In Software Product Lines (SPLs), variability information is an important part, which is not covered by traditional software metrics.
The SPL research community developed new variability-aware metrics to address this issue, which received increasing attention over the last decade \cite{El-SharkawyYamagishi-EichlerSchmid19, BezerraAndradeMonteiro+15, MontagudAbrahaoInsfran12}. 
In a previous study \cite{El-SharkawyYamagishi-EichlerSchmid19}, we identified 147 variability-aware metrics to measure qualitative characteristics of variability models and code artifacts, which partly influence each other \cite{BergerGuo14}.
While traditional software metrics for single systems are well analyzed with respect to their ability to draw qualitative conclusions \cite{RadjenovicHerickoTorkar+13}, there are only very few evaluations available regarding the application of variability-aware metrics for SPLs \cite{El-SharkawyYamagishi-EichlerSchmid19}. Further, there are no comparisons between well-established single system and variability-aware metrics available. The lack of available tools for measuring variability-aware metrics aggravates the situation.

In this paper, we present a concept for efficiently parsing code files of SPLs that stores sufficient information for the realization of single system metrics from traditional software engineering as well as variability-aware code metrics designed for the needs of SPLs. In addition, our concept allows the arbitrary combination of variability-aware code metrics with feature metrics, which was not investigated so far. 
Thus, the presented parsing concept provides the foundation for the realization and evaluation of new SPL metric suites like MetricHaven\footnote{\label{fn:MetricHaven}Available at \url{https://github.com/KernelHaven/MetricHaven}}. Here, we present the concepts behind the tool, which was presented in \cite{El-SharkawyKrafczykSchmid19}.
We pursue the following research questions:

\begin{enumerate}[label={\RQLayerOneLabel},leftmargin=*]
	\item \label{rq:Classical vs SPL Metrics} What are the requirements to support a flexible measurement of single system and variability-aware code metrics?
	\item \label{rq:Combination of Metrics} How can existing variability-aware metrics for code and variability models be combined?
	\item \label{rq:Efficient} What abstraction is required to support a scalable analysis of large-scale SPLs?
\end{enumerate}

We implemented our concept in the publicly available tool MetricHaven \cite{El-SharkawyKrafczykSchmid19}, which provides practitioners and researchers with a foundation for the flexible definition and measurement of code metrics for SPLs implemented in C. MetricHaven is also designed as a highly configurable software product line and provides re-implementations of traditional and variability-aware code metrics from different research groups. Its design supports the highly efficient measurement of more than 42,000 metric combinations on large-scale product lines.

Overall, we make the following contributions: 
\begin{itemize}
	\item We present the concept of Reduced Abstract Syntax Trees (RASTs) that contain sufficient information for the definition of most traditional and variability-aware code metrics, while minimizing resource overhead.
	\item A concept that allows a flexible combination of variability-aware feature and code metrics.
	\item A discussion of the limitations of the presented approach.
\end{itemize}

%% file: 020-relatedWork.tex
\section{Related Work}
\label{sec:Related Work}
The research community developed a huge variety of variability-aware metrics, designed for the needs of SPLs \cite{El-SharkawyYamagishi-EichlerSchmid19, BezerraAndradeMonteiro+15, MontagudAbrahaoInsfran12}. Below, we discuss the related work on variability-aware metrics based on four characteristics: \textit{Tool support}, \textit{applicability}, \textit{flexibility}, and \textit{scalability}.

\textbf{Tool support.} In 2012, Montagud et al.\ \cite{MontagudAbrahaoInsfran12} investigated to which extend authors of variability-aware metrics provide tool-support. Their study included metrics for all life cycles of SPLs and, thus, was not limited to implementation. They conclude that only 52\% of 35 identified papers provide (partial) tool support for the computation of metrics.
We address this issue by providing a concept together with a publicly available tooling for the flexible realization of variability-aware code metrics. The presented approach supports a broad variety of single system as well as variability-aware code metrics of different research groups \cite{El-SharkawyKrafczykSchmid19}.

\textbf{Applicability.} An important aspect is the applicability of the available metrics. We categorized implementation-related metrics according to four categories \cite{El-SharkawyYamagishi-EichlerSchmid19}: Metrics for \textit{variability models} (this was included, because variability models are used to manipulate all artifacts of SPLs), \textit{annotation-based code}, \textit{composition-based code}, and the combination of \textit{code and variability model metrics}. We discovered that available concepts and their realizations are limited either to one of the aforementioned categories or are further restricted to certain file types. For instance, \texttt{S.P.L.O.T.}\ \cite{MendoncaBrancoCowan09} and \texttt{DyMMer} \cite{BezerraBarbosaFreires+16} provide various metrics for variability models saved in the \texttt{S.P.L.O.T.}\ file format (XML files). \texttt{FEATUREVISU} \cite{ApelBeyer11} was used for the measurement of code artifacts from composition-based SPLs, using different feature-oriented implementation techniques. In the context of annotation-based code, many authors implemented their metrics to operate directly on the XML output of \srcml\footnote{\label{note:srcML}\url{https://www.srcml.org/}} \cite{LiebigApelLengauer+10, HunsenZhangSiegmund+16}. Thus, their measurement is limited to a specific set of implementation languages and require a re-implementation for the measurement of SPLs using a different annotation technique. Passos et al.\ \cite{PassosQueirozMukelabai+18} do not specify an implementation for the measurement of scattering degree metrics, but their appendix\footnote{\url{https://github.com/Mukelabai/featurescattering18/}} contains a set of \texttt{Bash} scripts explicitly designed for the analysis of Linux. This approach requires a re-implementation of their metrics for the measuring of other SPLs, even if they use a similar implementation technique.
We present a measurement concept for the analysis of annotation-based code artifacts of SPLs. In our implementation we decoupled parsing, data model, and the metrics computation from each other. Consequently, only a new parser is required for the analysis of SPLs realized with different programming languages.

\textbf{Flexibility.} Even if the variability model is often used for the configuration of code artifacts \cite{CzarneckiGruenbacherRabiser+12}, there are very few metrics available that include the complexity of the variability model when measuring code artifacts \cite{El-SharkawyYamagishi-EichlerSchmid19}. More precisely, we know only one study providing an evaluation for such a measure \cite{KolesnikovRothApel14}. Further, we do not know any comparisons of variability-aware code metrics with traditional metrics for single system metrics.
We present a concept that allows measuring of traditional and variability-aware code metrics in a single pass. For the use of variability-aware code metrics, we further allow the flexible integration of feature metrics to consider the complexity of the variability model.

\textbf{Scalability.} According to \cite{El-SharkawyYamagishi-EichlerSchmid19}, only 36\% of published metrics have been evaluated whether these metrics are sufficient to draw any qualitative conclusions. While some metrics have been applied on large-scale product lines from industry or publicly available SPLs, we did not discover any detailed examination of their runtime in general.
Our concept stores the information required for measuring different code metrics. We demonstrate the scalability of our approach by the application of 29,976 different metric variations on the Linux Kernel with more than 20,356 code files resulting in 53 GiB of measurement data. This is the first published performance analysis of SPLs metrics to the best of our knowledge.

%% file: 030-Tradeoffs.tex
\section[Tradeoffs in Designing Static Analysis Tools]{Tradeoffs in Designing Static\\Analysis Tools}
\label{sec:Tradeoffs}
\begin{figure}[tb]
	\centering
	\includegraphics[trim={0cm 4.5cm 10.5cm 0cm}, width=.8\columnwidth]{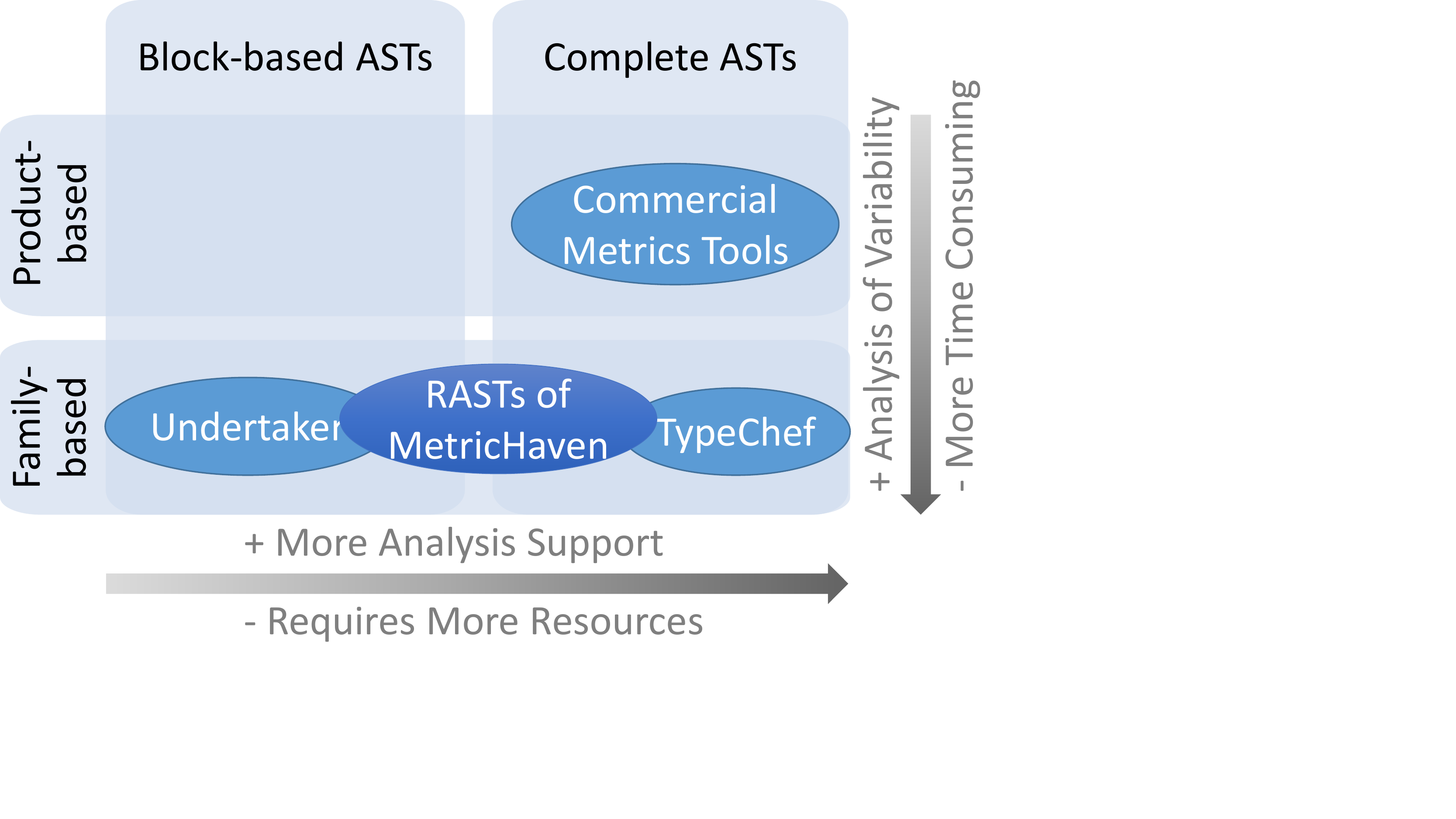}
	\Description{Categorization of static analysis approaches with respect to the used parsing approach.}
	\caption{Categorization of static analysis approaches with respect to the used parsing approach.}
	\label{fig:Categorization}
\end{figure}

Different parsing approaches exist for the static analysis of software, which result in different forms of  Abstract Syntax Trees (ASTs). These parsing approaches come with different tradeoffs. In the context of SPLs, there also exist different analysis strategies: Product-based, family-based, and feature-based analysis approaches~\cite{ThumApelKastner+14}. Below we discuss \mbox{(dis-)}advantages of these concepts and show why we choose a partial parsing approach in combination with a family-based analysis technique. Figure~\ref{fig:Categorization} provides an overview of the considered analysis strategies and parsing approaches together with a classification of our approach and existing analysis tools.

\input{041b-TableOfElements}

\subsection{SPL Analysis Strategies}
\label{sec:SPL Strategies}
Thüm et al.\ \cite{ThumApelKastner+14} surveyed analysis approaches for SPLs and identified three categories of analysis strategies:

\textit{Product-based analysis techniques} operate on instantiated products of the SPL. This strategy allows the usage of standard analysis techniques from traditional software engineering, since the variability information is resolved \cite{ThumApelKastner+14}. For instance, professional metric tool suites like the Axivion Bauhaus Suite\footnote{\url{https://www.axivion.com/en/products-60\#produkte_bauhaussuite}}, Teamscale from CQSE\footnote{\url{https://www.cqse.eu/en/products/teamscale/landing/}}, and SonarQube\footnote{\url{https://www.sonarqube.org/}} may be utilized for the measurement of instantiated code artifacts. However, for a high coverage of the original SPL, this strategy requires redundant computations as the products share code and, thus, is very time-consuming. Further, the analysis of all supported product variants of the SPL is often not feasible in practice as the number of products is typically exponential in the number of features.

\textit{Family-based analysis techniques} operate on product line artifacts containing variability information and take advantage of a variability model to limit the analysis to valid configurations only. This strategy allows analysis of the code for all possible product configurations, without the need of generating any products. However, this strategy does not work with available tools developed for the analysis of single systems. Since family-based analysis techniques consider all product line artifacts as a whole, the size of the analysis problem can easily exceed physical boundaries such as the available memory \cite{ThumApelKastner+14}.

\textit{Feature-based analysis techniques} analyze product line artifacts containing variability information, too. Contrary to family-based approaches, this strategy analyzes each feature in isolation and ignores all other features as well as the variability model. This reduces the potentially exponential number of analysis tasks. However, this kind of analyses cannot detect any problems caused by feature interactions \cite{ThumApelKastner+14}.

Most of the surveyed variability-aware metrics operate on product line artifacts containing variability information and consider all features, but ignore the variability model \cite{El-SharkawyYamagishi-EichlerSchmid19}. Thus, they can be classified somewhere in between family-based and feature-based analysis approaches. We designed our analysis approach so that it can reproduce the current state-of-the-art in variability-aware metrics but may also incorporate information from the variability model.

\subsection{AST Parsing Strategies for SPL Analyses}
\label{sec:AST Strategies}
We observed two fundamentally different parsing strategies for family-based analysis approaches. Sincero et al.\ \cite{SinceroTartlerLohmann+10} focus on parsing only \textit{preprocessor blocks} to extract variability information of product line artifacts. This approach takes advantage of the strong abstraction and allows the extraction of variability information in $\mathcal{O}(n)$ with the number of variation points. According to \cite{SinceroTartlerLohmann+10}, Undertaker\footnote{\url{https://vamos.informatik.uni-erlangen.de/trac/undertaker}} requires about half an hour to parse all 25,844 source code files (*.c, *.h, *.S) of the Linux Kernel Version 2.6.33 with a quad core CPU and 8 GB RAM. While this strategy is very fast compared to more detailed data representations, the analysis capabilities of this approach are very limited. The authors designed this approach for the analysis of (un-)dead code with respect to the implemented variability \cite{TartlerLohmannSincero+11}. This approach does not support any code analysis, since the parser does not parse any elements of the programming language.

\begin{figure*}[bt]
	\centering
		\includegraphics[width=0.95\textwidth]{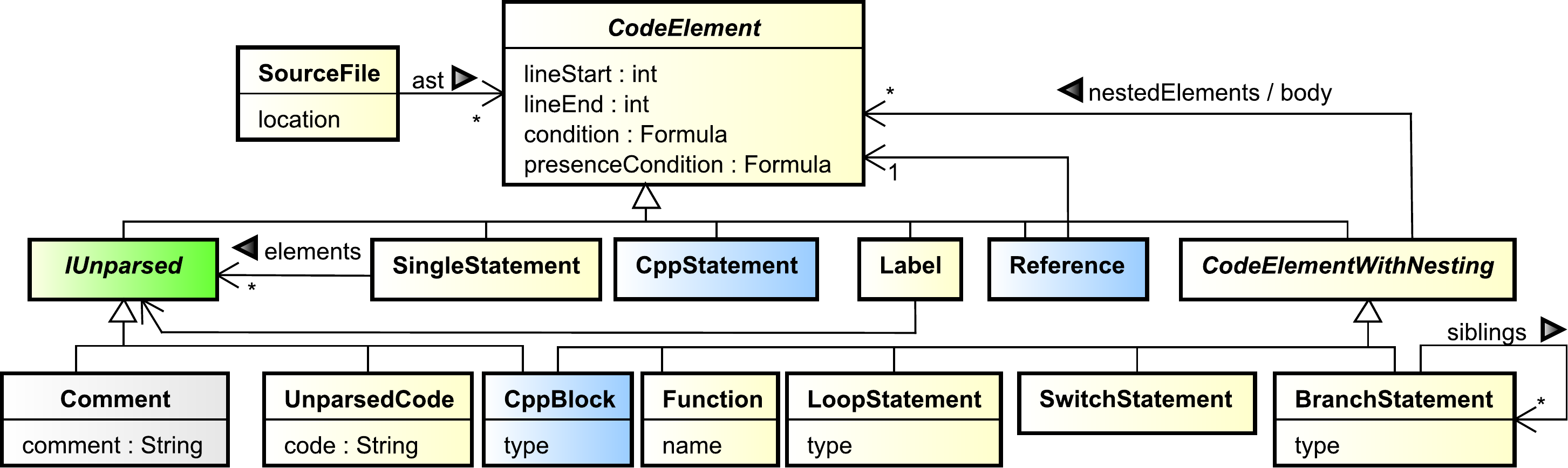}
		\vspace*{-10pt}
		\Description{Class diagram of presented RAST contains elements of annotation language (e.g., C-preprocessor statements) and programming language (e.g., C language) in one data model.}
		\caption{Simplified class structure used for parsing single system and variability-aware metrics (yellow:\ related to syntax elements of the programming language, blue:\ elements of the annotation language, green:\ related to both languages).}
		\label{fig:ClassDiagram}\vspace*{-10pt}
\end{figure*}

Kästner et al.\ \cite{KastnerGiarrussoRendel+11} use a more sophisticated parsing strategy consisting of a \textit{variability-aware lexer} and a \textit{variability-aware parser}, implemented as part of  TypeChef\footnote{\url{https://ckaestne.github.io/TypeChef/}}. The  lexer annotates all tokens of the programming language with its presence conditions, i.e., the condition of the enclosing variation point used for the selection of the token. It also  includes  all  header files and expands macros. The  parser creates for each supported configuration of the parsed code an alternative subtree as part of the resulting \textit{variable AST}. The authors use a SAT-solver during lexing and parsing to reason about code parts that belong together or may be skipped. The very detailed code representation in conjunction with annotated variability information allows a broad range of family-based analysis techniques, like variability-aware type checking, variable control-flow graphs, and variability-aware liveness analysis \cite{LiebigRheinKastner+13}. 
The creation of the very detailed variable AST requires much more effort than the previous approach. Parsing of the x86 architecture of the Linux Kernel version 2.6.33.3 with 7,665 C-files (*.h are included through the variability-aware lexer) requires roughly 85 hours on dual/quad-core lap computers with 2 to 8 GB RAM (the authors do not precisely specify their measurement system) \cite{KastnerGiarrussoRendel+11}. This parsing approach has an additional downside beside the massive time consumption. Through the macro expansion and the treatment of statements belonging to different configurations, the variable AST does not represent the developers view on the code anymore.

We surveyed existing traditional and variability-aware code metrics in order to design a Reduced AST. On the one hand, our RAST 
contains more information than the approach by Sincero et al., which stores only information about the variation points used in code artifacts. On the other hand, our approach stores less information than the variable AST and, thus, does not facilitate the same code analyses as supported by the TypeChef infrastructure. However, our concept provides an efficient measurement of a large variety of traditional and variability-aware code metrics, which can not be done by any of the previously discussed parsing strategies.

%% file: 041b-TableOfElements.tex
\begin{table*}[tb] \centering
  {\small
	 \renewcommand{\arraystretch}{.9}
    \begin{tabular}{@{} |l|l|p{3.7cm}|p{4.45cm}|p{4.9cm}| @{}}
      \hline
      \multicolumn{2}{|c|}{\textbf{Code Element}} & \textbf{No Variability} & \textbf{Variation Points (VPs)} & \textbf{Variation Point Expressions}\\
      \hline
			Only Variability&\cMark&\multicolumn{1}{c|}{---}&No.\ of {VP}s \cite{FerreiraMalikKaestner+16}, Cyclomatic Complexity on {VP}s \cite{Lopez-HerrejonTrujillo08,FenskeSchulzeMeyer+15}, Nesting Depth of {VP}s \cite{LiebigApelLengauer+10, ZhangBecker12, JbaraFeitelson13, HunsenZhangSiegmund+16}&\SDvp \cite{LiebigApelLengauer+10,CoutoValenteFigueiredo11,JbaraFeitelson13, PassosQueirozMukelabai+18,HunsenZhangSiegmund+16}, \SDfile \cite{ZhangBeckerPatzke+13, HunsenZhangSiegmund+16}, TD \cite{LiebigApelLengauer+10, JbaraFeitelson13, HunsenZhangSiegmund+16}\\
			\hline
			Function Definitions&\cMark&\multirow{2}{*}{Fan-In / Fan-Out \cite{Henry79}}&\multirow{2}{*}{Conditional Fan-In / Fan-Out}&\multirow{2}{*}{Degree Centrality \cite{FerreiraMalikKaestner+16}}\\
			\cline{1-2}
			Function Calls&\partialSupport&&&\\
			\hline
			Control Structures&\cMark&McCabe \cite{McCabe76}, Nesting Depth \cite{ConteDunsmoreShen86}&&\\
			\hline
			Statements&\cMark&Statement Count \cite{Jones86}&&\\
			\hline
			Line Numbers&\cMark&Lines of Code \cite{Jones86}&LoF \cite{LiebigApelLengauer+10, CoutoValenteFigueiredo11, ZhangBecker12, GaiaFerreiraFigueiredo+14, FenskeSchulzeMeyer+15, HunsenZhangSiegmund+16}, P{L}oF \cite{FenskeSchulzeMeyer+15, HunsenZhangSiegmund+16}&\\
			\hline
			Comments&\cMark&(Non-)Commented LoC \cite{Jones86}, Ratio of Comments per LoC \cite{FentonBieman14}&&\\
			\hline
			Operators \& Operands&\noSupport&Halstead \cite{Halstead77}&&\\
			\hline
			Variable Usage&\noSupport&Liveness of Variables \cite{ConteDunsmoreShen86}&&\\
			\hline
		\end{tabular}	
  }
  \caption{Supported measures (code elements $\times$ variability dimensions; \cMark\ = represented in RAST, \partialSupport\ = supported only via String operations, \noSupport\ = no support).}
  \label{tab:Supported Metrics}\vspace*{-24pt}
\end{table*}

%% file: 040-concept.tex
\section{Concept}
\label{sec:Concept}
Here, we present the concept of parsing Reduced Abstract Syntax Trees (RASTs). This was motivated by designing a tailored parsing approach which is able to extract the information needed for the desired static analyses. In our case, we planned a flexible definition of single system and variability-aware code metrics to allow comparisons of them. Based on our survey \cite{El-SharkawyYamagishi-EichlerSchmid19} on variability-aware code metrics and an informal literature study on metrics from traditional software engineering, we came up with the following requirements for parsing RASTs (cf.\ \ref{rq:Classical vs SPL Metrics}):

\begin{enumerate}[label={\ReqLabel},leftmargin=*]\label{req:list}
	\item \textit{Parsing of un-preprocessed code}. While established metric analysis tools from commercial vendors usually resolve preprocessor statements before conducting metrics, variability-aware metrics analyze the preprocessor statements. Thus, we require a common data representation for code annotations (in our case C-preprocessor statements) and for elements of the programming language (in our case AST elements of the C-language). This is a challenging task, since the used preprocessor is not part of the programming language and can be used at arbitrary positions inside a code file, independently of any syntax definitions.
	
	\item \label{rq:No syntactically correct AST}\textit{No syntactically correct AST needed}. An important aspect is to which extent the resulting AST-structure needs to support only syntactically correct programs. Contrary to compilation tasks and type checking analyses, we do not need a syntactical correct AST for the computation of code metrics. However, the AST structure should be as close as possible to the actual code structure to simplify the definition of code metrics. Thus, it is still a challenging task to enhance a traditional AST structure with variability annotations, since these annotations may be inserted at arbitrary positions intertwined with AST elements of the programming language.
	
	\item \textit{Granularity of RAST}. For optimization as well as for practical reasons it is important to assess the required granularity of parsed elements. A very fine grained AST, containing representations for all syntax elements of the annotation and programming language, conceptually supports every code metric. On the other hand, this requires much more effort to develop a very comprehensive parsing approach and leads to higher resource consumption. Due to limited development resources, we designed a Reduced Abstract Syntax Tree (RAST), which is sufficient for measuring all planed metrics and may be easily extended to support further metrics, if desired. The granularity of the RAST is driven by the measured elements of surveyed metrics, which we present in Table~\ref{tab:Supported Metrics}.
\end{enumerate}

%% file: 042-RAST.tex
\vspace*{-7pt}
\subsection{Reduced Abstract Syntax Tree (RAST)}
\label{sec:Concept:AST}
Based on our SLR on variability-aware code metrics \cite{El-SharkawyYamagishi-EichlerSchmid19} and an informal literature study on metrics for single systems, we designed a Reduced Abstract Syntax Tree (RAST) for the efficient measurement of the most relevant traditional and variability-aware code metrics. Our goal is the measurement of C-based SPL implementations.\footnote{The  concepts we propose here could also be applied well beyond C.}
Thus, we limited the scope of our analysis to the measurement of metrics on a per-function basis. Figure~\ref{fig:ClassDiagram} presents the main elements of our RAST:

\input{044-controlflow}

\begin{itemize}
	\item \texttt{SourceFile}s represent the RAST representation of code files.
	
	\item The \texttt{CodeElement} is the super class of all RAST elements. It stores the line numbers to trace parsed elements back to their location in code files and facilitates LoC-metrics. Further, we store for each element two representations of the condition of surrounding variation points: The \texttt{condition} stores the condition of the  innermost variation block, considering conditions of siblings for \elif/\elseCPP-blocks. For instance, we store the condition \texttt{$A$} of the \texttt{while} statement in Line~\ref{lst:condWhile} of the listing in Figure~\ref{fig:Concept:ControlFlow}. This allows the computation of feature-based metrics on all parsed elements, e.g., \textit{Scattering Degree} metrics. Second, \texttt{presenceCondition} provides an alternative as it stores the full presence condition for the inclusion of the element, also considering all surrounding variation points. For code elements that are not surrounded by any variation points, we set \texttt{condition} and \texttt{presenceCondition} to \texttt{TRUE}. 
	
	\item The \texttt{SingleStatement} is the most fine-grained element of the RAST. We do not provide RAST representations for expressions of statements, but we store these elements as unstructured text (\texttt{UnparsedCode}). For instance, in Line~\ref{lst:stmt} we store a \texttt{SingleStatement} with the text ``\texttt{stmt;}'', not knowing whether this is a function call, a variable declaration, or anything else.
	
	\item The \texttt{IUnparsed} element facilitates the storage of preprocessor elements at arbitrary positions inside the RAST. This is required since preprocessor directives are not syntactical elements of the C programming language and may be used at arbitrary positions inside a code file, independently of any syntax definitions. The \texttt{IUnparsed} element is the parent of all (parsed) \texttt{CppBlock}s and \texttt{UnparsedCode} expressions of \texttt{SingleStatement}s.
	
	\item We use \texttt{CppBlock} to store conditional blocks, i.e., variation points. This means, we store \texttt{\#if}, \texttt{\#ifdef}, \texttt{\#ifndef}, \texttt{\#elif}, and \texttt{\#else}-blocks in separate instances, referring to all siblings of the same block structure. The \texttt{type} attribute is used to distinguish between the different preprocessor elements and to allow a differentiation during the computation of metrics, if required. \texttt{CppBlock} inherits from \texttt{IUnparsed}, which is used for elements of \texttt{SingleStatement}s, and inherits from \texttt{CodeElementWithNesting}, which is used as a container inside our RAST. The multiple inheritance allows a nesting of preprocessor directives at arbitrary positions inside the RAST.
	
	\item We use \texttt{BranchStatement}s similar to \texttt{CppBlock}s to store the \texttt{if} and \texttt{else} statements of the programming language. This class also stores the siblings of the same if/else-structure. Again, the \texttt{type} denotes which specific syntax element was used to allow a differentiation during the metrics computation, if necessary.
	
	\item \texttt{LoopStatement}s represent any loop of the programming language. Contrary to \texttt{BranchStatement}s they do not have siblings. Again, we support different loop \texttt{type}s.
	
	\item \texttt{Function}s represent function definitions. The function's signature is stored as \texttt{UnparsedCode}, while the function body is composed of previously described elements.
	
	\item \texttt{Reference} elements are special as they neither represent syntax elements of the programming language nor of the annotation language. They are used in case that syntactical elements of the presented RAST, like loops or control structures, are split into multiple parts by C-preprocessor statements. The listing of Figure~\ref{fig:Concept:ControlFlow} shows an example in which the C-preprocessor is used for the conditional compilation of a loop statement, while the statements of the loop are always present. In this case, a \texttt{LoopStatement} with one \texttt{Reference} is stored inside a \texttt{CppBlock}. This can be seen on the right side of Figure~\ref{fig:Concept:ControlFlow}. The actual statements are stored outside of the \texttt{CppBlock}. This way it is possible to simultaneously define metrics on the same parsed data structure, that consider the nested statements as variable as well as metrics that do not treat this statement as variable.
\end{itemize}

%% file: 044-controlflow.tex
\begin{figure*}
  \centering
	\begin{minipage}{.215\textwidth}
		\includegraphics[trim={0cm 12.5cm 25.8cm 0cm},clip,scale=.4]{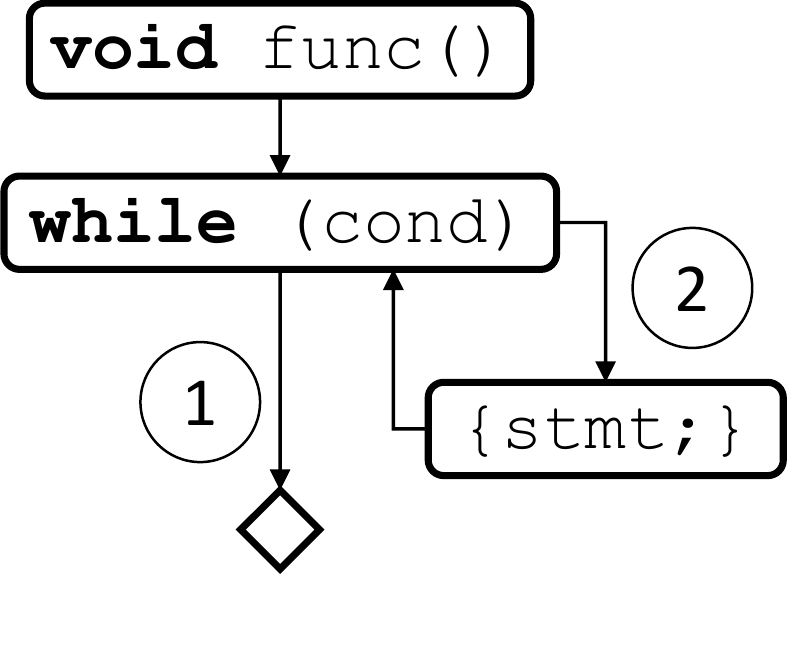}
	\end{minipage}
	\begin{minipage}{.255\textwidth}
\begin{lstlisting}[language=cWithPre]
 void func() {
   #ifdef(A)%*\tRef{featureA}*)
     while (%*$cond$*)) %*\label{lst:condWhile}*)
   #endif 
     {
         stmt; %*\label{lst:stmt}*)
     }
 }
\end{lstlisting}
		
		\begin{tikzpicture}[remember picture, overlay,
				every edge/.append style = { ->, thin, >=stealth, black!30, dashed, line width = 1pt },
				text width = 2.75cm]

		\node [above right=.7cm and 0.6cm of featureA, text width=2.78cm, rotate=270, style = {align = center, minimum height = 10pt, font = \bfseries}](Features) {\small Feature Reference};
		\draw[thin] (Features.197) edge (featureA.west);
		\end{tikzpicture}
	\end{minipage}
	\begin{minipage}{.38\textwidth}
		\includegraphics[trim={0cm 10.5cm 14.1cm 0cm},clip,scale=.4]{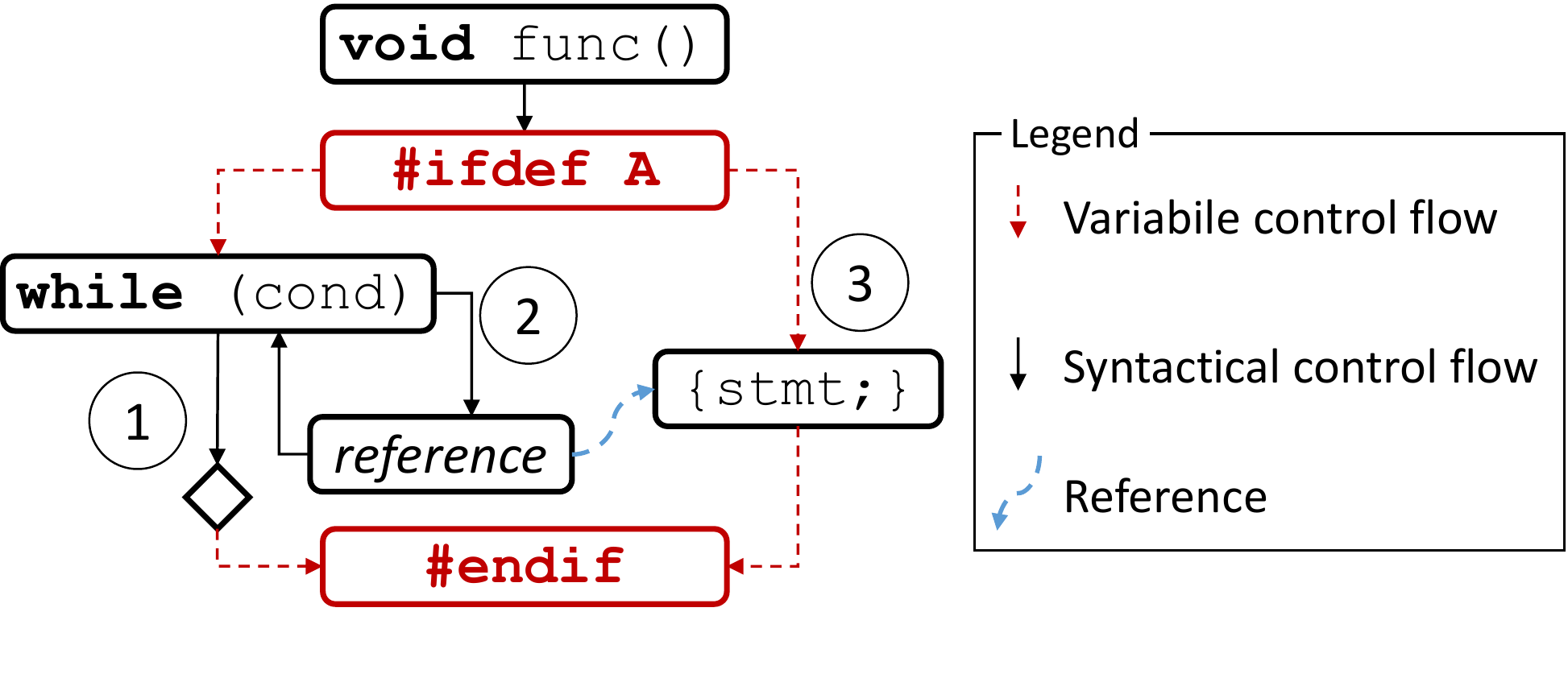}
	\end{minipage}
  \vspace*{-5ex}
	\Description{This figures presents a fictive example of a C-code function. The function contains a while-loop with a nested statement. An ifdef-statement surrounds the while-statement, but not the nested statement. Thus, iterative part of the loop is conditional, while the remainder of the code is always present in the code function. The left part shows a control flow, while ignoring the variability. This control flow contains the while statement. The right side contains a variable control flow, which covers also the preprocessor statements.}
	\caption{Code snippet and its (variable) control flow representations as it can be measured by MetricHaven.}
	\label{fig:Concept:ControlFlow}
	\vspace*{-3ex}
\end{figure*}
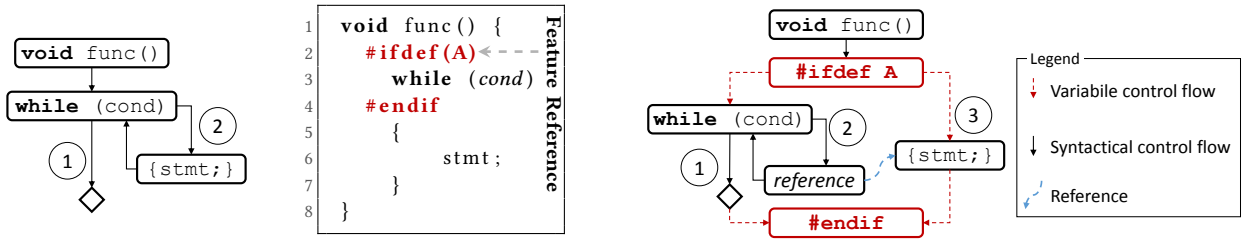

%% file: 043-application.tex
\begin{figure*}[tb]
	\centering
	\includegraphics[trim={0cm 10cm 4.0cm 0cm},width=0.7\textwidth]{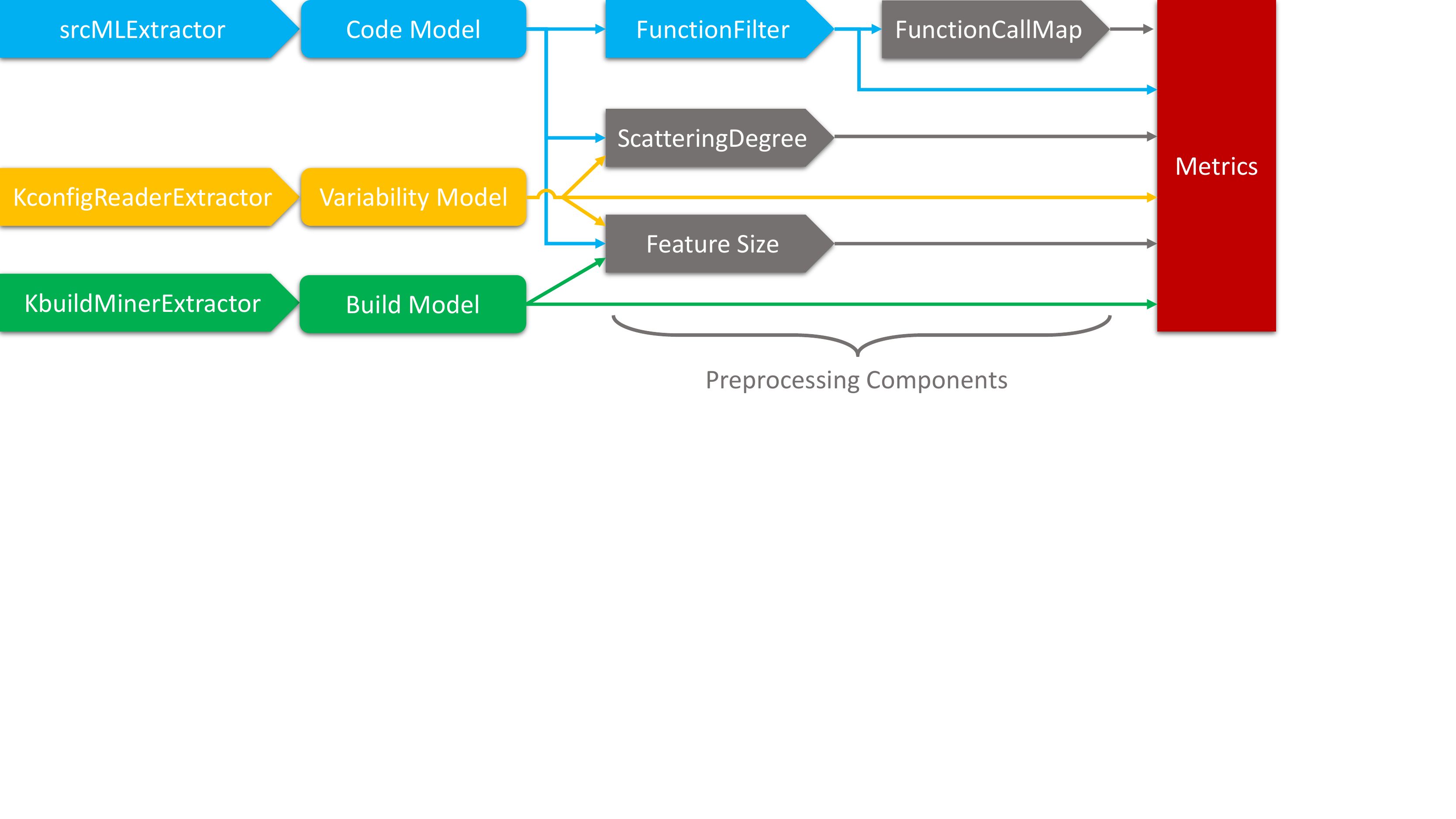}
	\vspace*{-2ex}
	\Description{Analysis pipeline of MetricHaven. Three pipelines run in parallel starting with parsing of code, build, and variability models. This information is stored in three models. Preprocessing elements use the three models to compute information required for the metrics. These are the FunctioCallMap, the ScatteringDegree, and the FeatureSize components. The information of the preprocessing components as well as the data models of the parsing components are used as input for the metrics computation.}
	\caption{Analysis pipeline structure used in MetricHaven.}
	\label{fig:MetricHavenPipeline}
	\vspace*{-3ex}
\end{figure*}

\vspace{-10pt}
\subsection{Application of RAST}
\label{sec:Concept:Application}
The RAST is designed to preserve the code structure in order to facilitate the computation of variability-aware code metrics according to their original definitions, while reducing the performance overhead by omitting elements that are not required for the metrics. 
For this we neither resolve the variability as usually done by commercial metric tool suites nor do we duplicate parsed code elements of alternative variants as done by TypeChef \cite{KastnerGiarrussoRendel+11}, as this would lead to modified metric values.
As a consequence, our RAST contains a 150\% representation of the parsed code.

Parsing of C-code requires the ability to cope with \textit{undisciplined annotations} \cite{LiebigKastnerApel11, MedeirosRibeiroGheyi+18}, which are conditional compilation directives that do not align with the underlying syntactic structure of the code.
Our RAST provides two concepts to support these annotations as described above:
In cases that elements of a statement are conditional, a \texttt{CppBlock} inside a \texttt{SingleStatement} may be used to store the conditional elements.
Further, \texttt{Reference}s may be used to represent conditional control structures.

Based on the RAST, new metrics may be implemented as a visitor,\footnote{According to the visitor design pattern.} which may be parameterized to represent different variations of a metric family.
For instance, MetricHaven uses one visitor to compute different variations of McCabe's Cyclomatic Complexity measure \cite{McCabe76}, which counts the linear independent paths of the (variable) control graph.
For the single system version, we completely ignore the variability of the code and add 1 to the number of visited control structures (\texttt{while}, \texttt{for}, \texttt{if}, \texttt{case}).
According to \cite{ConteDunsmoreShen86} this counting approach is equivalent to the original definition of McCabe.
However, this approach provides support for control structures of undisciplined annotations, since we do not need to compute a syntactically correct control graph.
The left side of Figure~\ref{fig:Concept:ControlFlow} provides an example on how we compute the cyclomatic complexity of a conditional loop.
By ignoring the annotations, we detect two linear independent paths of the resulting control graph.
The variability-aware version of this metric considers only paths created by variation points \cite{Lopez-HerrejonTrujillo08}, i.e., \texttt{\#ifdef}-statements.
Finally, we provide a superimposition of both variants by counting the number of control structures of the programming language and the annotation language.
The right side of Figure~\ref{fig:Concept:ControlFlow} visualizes the resulting control graph, which contains three linear independent paths:
The loop may be present but omitted completely at run-time \textcircled{1}, the loop may be executed \textcircled{2}, and the loop may be removed through conditional compilation but the statements are kept \textcircled{3}.

%% file: 050-realization.tex
\section{Realization}
\label{sec:Realization}

Our concept is implemented as a prototype for the analysis of C-based SPLs like the Linux Kernel. This is realized in several plug-ins for the \kernelhaven\ infrastructure \cite{KroeherEl-SharkawySchmid18a,KroeherEl-SharkawySchmid18b}. This infrastructure supports three types of extractors to read information from the product line to be analyzed:
\begin{itemize}
	\item Code extractors extract variability information from the source code implementing the software product line. For C source code, this typically involves parsing \ifDefBlock{}s of the C-preprocessor.
	\item Build model extractors extract variability information from the build process of the product line. This typically involves presence conditions that define in which configurations a given source code file is compiled into the product line.
	\item Variability model extractors extract the variability model of the software product line. This contains a list of all features and constraints between them.
\end{itemize}
In \kernelhaven, the result from these extractors is represented in models that are agnostic to implementation details of specific product lines, while still being extensible to additional information. This allows the following analysis components, which access these models, to be implemented independently from implementation details of specific product lines. The following paragraphs will first introduce which extractors were used in our prototype implementation and then explain the analysis process that implements our approach.

The parsing of the product line source code is implemented in the \srcmlextractor\footnote{\url{https://github.com/KernelHaven/srcMLExtractor}} plug-in for \kernelhaven. It is based on the \srcml\textsuperscript{\ref{note:srcML}} tool which parses source code to an XML format \cite{collard2011lightweight}. The \srcmlextractor{} parses this XML and converts it into a model compatible with \kernelhaven. The extension mechanism of the general code model in \kernelhaven{} is used to model the Reduced Abstract Syntax Tree (RAST), as introduced in Section~\ref{sec:Concept:AST}. The parsing process of the \srcmlextractor{} does not parse the full AST output of \srcml, but only descends to a granularity that provides sufficient information to build the RAST. See Figure~\ref{fig:ClassDiagram} for a simplified class diagram of the resulting RAST structure.

For the build and variability models, we use the \kbuildminerextractor\footnote{\url{https://github.com/KernelHaven/KbuildMinerExtractor}} and \kconfigreaderextractor\footnote{\url{https://github.com/KernelHaven/KconfigReaderExtractor}}. 
The former extracts variability information from the \texttt{Kbuild} build process of the Linux Kernel, the latter reads the \texttt{Kconfig} variability model present in the Linux Kernel source tree. These plug-ins and their underlying tools were already used in previous analyses of the Linux Kernel, and there were no changes done to these plug-ins when implementing the approach presented in this paper.

The analysis process in \kernelhaven{} typically consists of multiple analysis components that are combined to an analysis pipeline. The output of the previous component(s) is used as the input for the following component(s). The initial input for the first analysis component(s) are the models supplied by the three extractors (see above). This structure allows for simple re-use of analysis components when creating new analysis pipelines.

The calculation of metrics is implemented as such an analysis pipeline in the \metrichaven$^{\ref{fn:MetricHaven}}$ plug-in for \kernelhaven. Figure~\ref{fig:MetricHavenPipeline} shows an overview of this pipeline structure. The coloring of the lines indicate the flow of the three models extracted from the product line, as described above. The actual metric computation happens in the rightmost component at the end of the pipeline. The input for this component are the three extracted models (the code model went through the \texttt{FunctionFilter} first) and the output of three preprocessing components.

\begin{itemize}
	\item The \texttt{FunctionFilter} component splits the code model into individual functions, and removes any elements that are outside of functions (such as global variables). This component does not compute any values for metrics; it is only used for convenient data organization. The result is a stream of code functions.
	\item The \texttt{FunctionCallMap} component analyses calls between functions. Since the Reduced Abstract Syntax Tree (RAST) is not fully parsed (cf.~Section~\ref{sec:Concept:AST}), function calls inside statements are identified heuristically: If the unparsed code string of a statement contains a function name followed by an opening parenthesis, we consider that statement to contain a call to this function. For each identified function call, the calling function (caller), the called function (callee), and the presence condition and location of the statement containing the function call are stored. This information is for example used in the Fan-In/-Out and Degree Centrality metrics.
	\item The \texttt{ScatteringDegree} component calculates the Scattering Degree metric for all features of the variability model. The result is a map of all features and their values for the different scattering degree types (\SDvp\ \cite{LiebigApelLengauer+10,CoutoValenteFigueiredo11,JbaraFeitelson13, PassosQueirozMukelabai+18} and \SDfile \cite{ZhangBeckerPatzke+13, HunsenZhangSiegmund+16}).
	\item The \texttt{FeatureSize} component calculates the Feature Size metric for all features in the variability model. The result is a map of all features and the number of statements controlled by the feature.
\end{itemize}

The preprocessing components are executed before the final metric computation component, because they require a full overview of the complete code model. In contrast, the metric calculation component calculates the metric values on a per-function basis. This reduced view on the code model allows to reduce the complexity of the metric calculation component and also helps to mitigate a memory problem. Since our implementation scales to a large number of metrics to be calculated per function, the amount of resulting metric values can grow quickly. In practice, the memory required to store this are several gigabytes. With the per-function approach in the metric calculation component the results of a single function can directly be written to disk, freeing the main memory.

Our implementation offers a number of configuration options. Most importantly, it allows for free selection of metrics to calculate. The user can select anything from 1 up to 42,796 metrics and metric combinations to be calculated per function. Additionally, the extractors at the beginning of the pipeline can be exchanged. This enables our infrastructure to run on different software product lines, while the analysis components require no adaptation. This is because the models used to represent the extraction result are agnostic to specifics of single product lines. Finally, the number of threads used in the code extraction plug-in and the metric calculation component can be configured independently. See the evaluation in Section~\ref{sec:Evaluation} for details on the potential performance improvements.

%% file: 070-benchmark.tex
\section{Evaluation}
\label{sec:Evaluation}

We ran our prototype implementation on the x86 architecture of the Linux Kernel version 4.15\footnote{https://mirrors.edge.kernel.org/pub/linux/kernel/v4.x/linux-4.15.tar.xz} to evaluate the implementation of our concept. Based on the RAST presented in Figure~\ref{fig:ClassDiagram}, we were able to realize various single system and variability-aware code metrics \cite{El-SharkawyKrafczykSchmid19}. Through the combination of variability-aware code metrics with feature metrics we support 42,796 metric variations (as of Summer 2019), which can be measured in a single pass (cf.\ \ref{rq:Efficient}). While most metrics can be implemented in a straightforward manner, some implementations require adaptations for our RAST. For instance, the detection of function calls for the Fan-In/-Out metric can only be implemented heuristically, since it would require full parsing of the expression syntax (cf.\ Section~\ref{sec:Realization}).

The x86 architecture in the Linux Kernel has 20,356 C-source files that are evaluated by us.
106 ($\approx 0.5\%$) of these files, cannot be handled by the current implementation of the \srcmlextractor, which translates the XML output of \srcml{} to \kernelhaven's code model.
This is mostly related to very special corner cases of \textit{undisciplined} C-preprocessor usage \cite{MedeirosRibeiroGheyi+18}, i.e., conditional annotations that do not align with the syntactic structure of the code.
\srcml{} marks up the C-syntax independently of C-preprocessor directives.
In conjunction with \textit{undisciplined} C-preprocessor directives, this can lead to incorrect markups provided by \srcml{}.
Proper handling of those structures requires adaptations of the \srcml{} parser.
Some special cases are detected and fixed by our \srcmlextractor.
However, this approach cannot repair all of these cases and requires significant development effort.
Further, a minority of these corner cases cannot be mapped to our RAST at all as they violate the few structural assumptions of the RAST.
An even more lenient RAST that can model these cases, however, would significantly complicate the definition and computation of metrics.



We ran two sets of metrics on the Linux Kernel, to evaluate scalability of our approach (cf.\ \ref{rq:Efficient}): First, a selected subset of metric combinations that we also used in practice for our own work with the Linux Kernel. We call this set \textit{atomic metrics}. It contains all basic code metrics and all possible combinations of code metrics combined with a single feature metric. Code metrics combined with multiple feature metrics are not included. This results in a set of 648 metrics. Second, we allowed all metric combinations. However, the implementation of one metric family (approximation of Eigenvector Centrality) requires significantly more memory than the other implementations and is not optimized for the provided parallelization capabilities of MetricHaven. For this reason, we executed only the 148 metric variations of this metric, which were already executed as part of the atomic metrics, while we executed all variations of the remaining metrics. This results in a set of 29,976 metrics.

For the performance measurements, we ran our implementation in a virtual machine running Ubuntu 16.04 with 40 logical CPU cores of an Intel Xeon E5-2650v3 @ 2.3 GHz and 314 GiB RAM. For the analysis, we limited the JVM once to 50 GiB memory\footnote{via the command line switches: -Xmx50g -Xms50g} and once to 24 GiB. However, it must be noted that the extractors run in separate processes as they execute external tools and, thus, allocate additional memory. Accurate timings for specific phases are hard to measure since KernelHaven makes heavy use of parallelization. For example, the preprocessing components already start to run while code parsing is still running. However, the actual metric calculation component can only start when the complete code model has been passed through the preprocessing components. That means that we identify two distinct execution phases: code parsing and metric calculation. The preprocessing phase, which partially happens in parallel to the code parsing only takes a few seconds, which is insignificant compared to the total runtime. Thus, we do not supply measurements for this phase. The code parsing and metric calculation components can also be independently configured to use a specified amount of threads. We measured the runtime of these components for different numbers of configured threads.
For the experiments, we used a range of 1 to 10 threads to cover a spectrum which is supported by most workstation computers.

Running the 648 \textit{atomic metrics} on all 409,253 functions that we parse from the x86 Linux Kernel architecture produces 265,195,944 measures. In CSV format, this is about 1.2 GiB. The metric calculation step takes about 27.5 minutes (54 minutes with 25 GiB memory for the JVM) to run on a single thread and can be decreased to about 13.75 minutes (38.5 minutes) on 10 computation threads.

Running the large set of 29,976 metrics on all 409,253 functions that we parse from the x86 Linux Kernel architecture produces 12.2 billion measures.
In CSV format, this is about 53 GiB.
The metric calculation step takes from 11 hours and 26 minutes on one thread to 6 hours 15 minutes on 10 threads (6 hours 20 minutes total runtime).
The parsing performance stays the same as described in the previous paragraph as it is independent from the number of computed metrics.
The parallelization benefit of metric computations is slightly less pronounced compared to the \textit{atomic metrics}.
However, our prototype is not fully optimized with regard to parallelization.

\begin{figure}[tb]
	\centering
	\includegraphics[trim={0.04cm 8.5cm 20.95cm 0cm},clip,width=\columnwidth]{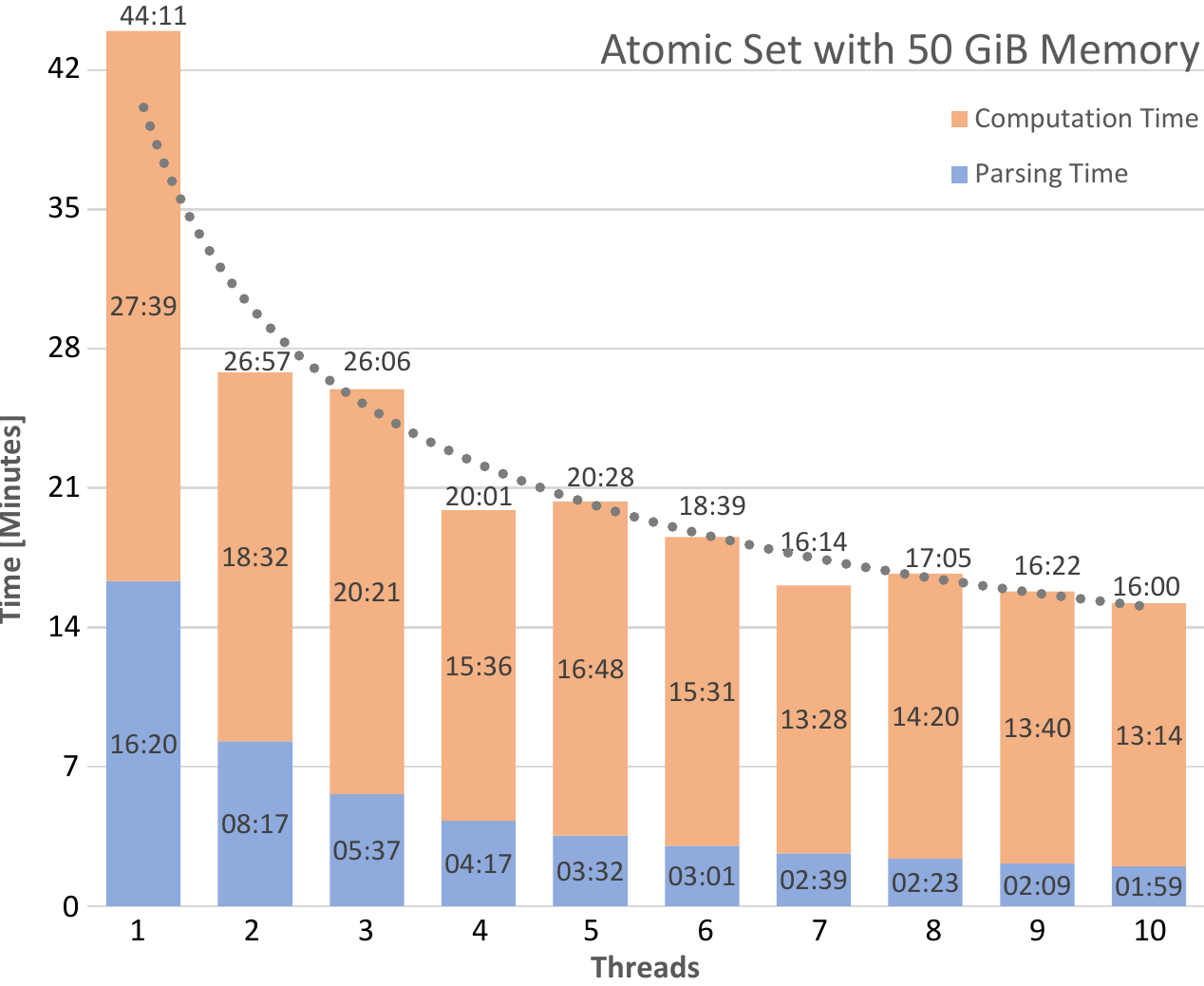}
	\Description{Benchmark shows performance of MetricHaven to measure 648 metric variations on the Linux Kernel.
	The parsing of code files takes from about 16 minutes and 20 seconds on 1 thread to about 2 minutes on 10 threads.
	The metrics runtime takes about 27 minutes and 39 seconds on 1 thread to about 13 minutes and 14 seconds on 10 threads.
	The total runtime takes about 44 minutes and 11 seconds and goes down to 16 minutes on 10 threads.}
	\vspace*{-18pt}
	\caption{Parsing/computation/total runtime of 648 \textit{atomic metrics} with different number of threads.}
	\label{fig:ExtractorTimings}
\end{figure}

Figure~\ref{fig:ExtractorTimings} visualizes the runtime of running the 648 \textit{atomic metrics} on a different number of threads with 50 GiB memory.
The number of threads for the parsing and the metric computation phase were both modified together on a range of 1 to 10.
For each run, we present the parsing time (lower, blue bars), the metrics computation time (upper, orange bars), and the total execution time (numbers above), which requires some additional time for loading the plug-ins and writing the results.
The gray, dotted line indicates the trend line of the total execution time.
Running the experiments with less memory results in a similar behavior.
However, the performance gain of configuring more threads is much smaller compared to the experiment with 50 GiB memory, because of the increased workload of the JVM garbage collector.
Please note that MetricHaven keeps the parsed RAST of the complete Linux Kernel in memory when  it starts its computation.

%% file: 060-discussion.tex
\section{Discussion} 
\label{sec:Discussion}

In the development of any kind of analysis tool, one needs to make a number of tradeoffs. These relate, in particular, to generality (what range of metrics to support), speed, and range of supported artifacts.
The key principles we used for developing the solution, described here, are: 
\begin{itemize}
\item \textit{Genericity:} It should support a large range of product line metrics. 
Basically, it should support all metrics that were documented as product line metrics so far and  at the same time, it should support a very large (though not necessary complete) range of single system metrics. 
\item \textit{Scalability:} It should be able to deal with very large product lines and very large numbers of metrics simultaneously.
\item \textit{Performance:} It should perform these tasks very efficiently. 
\end{itemize}

The key innovation was to introduce the solution of a reduced abstract syntax tree (RAST) and to tailor it very well to the task at hand. This avoids the representation of language details that are not relevant to the metrics analysis and abstracting those that are relevant as far as possible.  A detailed analysis of the representational needs of the metrics provided the basis of our design. 
At this point also some trade-offs had to be made.  As a result, we do not support all kinds of metrics. For example, we do not support Halstead metrics \cite{Halstead77}. 
%
%

In order to achieve high performance, we create the metrics values nearly completely in a single pass. This leads to significant performance improvements as technical properties like CPU caching are used in an optimal way and the generation of the data structures does not need to be made multiple times.
The exception is the approximation of Eigenvalue Centrality, which requires a two-pass approach.
 Overall, we managed to get an extremely high performance, using our approach along with very good scalability properties, both in terms of the number of metrics and analyzed code size. For example, in our evaluation we found 6 hours 20 minutes for analyzing the complete Linux Kernel (cf.\ Section~\ref{sec:Evaluation}), which means that about 0.76 seconds were required per metric. Or, to put it differently, for producing 29,976 metrics, we needed about 0.06 seconds per function. We regard this as an extremely strong performance, although it is very difficult to compare as there is no other metrics tool that supports a similarly wide range of metrics.

%% file: 080-conclusion.tex
\section{Conclusion} 
\label{sec:Conclusion}

In this paper, we presented the concept of MetricHaven, for simultaneously evaluating a large number of metrics on very large product lines in a highly efficient manner. MetricHaven supports more than 42,000 metrics, requiring less than 0.06 seconds for computing a selection of 29,976 metrics per function of the Linux Kernel leading to a total execution time of about 6 hours and 20 minutes for analyzing the whole Linux Kernel, yielding more than 53 GB of metrics data.
The approach is highly customizable (achieving significant speed-ups when reducing the number of metrics to process).
In particular, beyond analyzing product line metrics, it is also capable of creating a significant range of single system metrics. 

The key to achieving these capabilities, was first to identify key realization requirements as required in \ref{rq:Classical vs SPL Metrics}. In Section~\ref{sec:Concept}, we introduced three core requirements that enabled us to create this approach: we had to directly parse un-preprocessed code (1), yielding an AST which is not a syntactically correct representation (2). Further, we had to minimize the required information by abstracting the information to a significant extent, leading to a rather coarse-grained AST with reduced information (RAST) (3).

For \ref{rq:Combination of Metrics}, the answer was actually rather simple: by having an integrated representation that does not replicate basic code elements (as, for example, some approaches to handling variable code do \cite{KastnerGiarrussoRendel+11}) and integrating the variability given by the preprocessor information, we could simply handle the subset of non-variable information also from a metrics point of view. 

We addressed \ref{rq:Efficient} by creating the notion of a reduced abstract syntax tree (RAST). The abstraction level of this is tailored to exactly the level of detail required for handling all relevant product line metrics. All further information is skipped, respectively, not parsed in detail. We described this in detail in Section~\ref{sec:Concept:AST}.
 
In future work, we plan to further extend this framework in terms of the range of supported metrics, improve its performance and apply it to study numerous properties of product line implementations. We are particularly interested in the prediction of defects based on product line metrics.